\documentclass[twocolumn,preprintnumbers,superscriptaddress,nofootinbib,aps,prd,floatfix]{revtex4-2}

\usepackage[toc,page,header]{appendix}

\usepackage{titletoc}

\usepackage{epsf,epsfig,subfigure,graphicx,amsmath,amssymb,cancel}
\usepackage{subfigure}
\usepackage{soul}
\usepackage[colorlinks=true,urlcolor=purple,anchorcolor=blue,citecolor=blue,filecolor=blue,linkcolor=red,menucolor=blue]{hyperref}
\usepackage[usenames,dvipsnames]{xcolor}
\usepackage{slashed}
\usepackage{bm}
\usepackage{bbm}
\usepackage{enumitem}
\usepackage{myfuncs}
\usepackage{orcidlink}

\usepackage{afterpage}
\usepackage[utf8]{inputenc}
\usepackage{multirow}
\usepackage{makecell}

\allowdisplaybreaks

\newcommand{\newc}{\newcommand}
\newc{\gsim}{\lower.7ex\hbox{$\;\stackrel{\textstyle>}{\sim}\;$}}
\newc{\lsim}{\lower.7ex\hbox{$\;\stackrel{\textstyle<}{\sim}\;$}}
\newc{\gev}{\,{\rm GeV}}
\newc{\mev}{\,{\rm MeV}}
\newc{\ev}{\,{\rm eV}}
\newc{\kev}{\,{\rm keV}}
\newc{\tev}{\,{\rm TeV}}
\newc{\MHT}{$H_T^{\text{miss}}$}
\newc{\MET}{$\slashed{E}_T$}
\newc{\MTT}{$M_{T2}$}

\def\ln{\mathop{\rm ln}}

\newc{\mz}{M_Z}
\newc{\mpl}{M_*}
\newc{\mw}{m_{\rm weak}}
\newc{\nr}[1]{N^c_R{}_{#1}}

\newcommand{\Uone}{\text{U(1)}}

\definecolor{H_D_He_naDM}{HTML}{E69F00}
\definecolor{D_He_naDM}{HTML}{D55E00}
\definecolor{H_D_naDM}{HTML}{56B4E9}
\definecolor{D_naDM}{HTML}{0072B2}
\definecolor{H_naDM}{HTML}{CC79A7}
\definecolor{N_SIDM}{HTML}{009E73}

\def\beq{\begin{equation}}
\def\eeq{\end{equation}}
\newcommand{\bea}{\begin{eqnarray}\begin{aligned}}
\newcommand{\eea}{\end{aligned}\end{eqnarray}}
\def\bitem{\begin{itemize}}
\def\eitem{\end{itemize}}

\definecolor{darkgreen}{rgb}{0,0.6,0}
\definecolor{darkorange}{rgb}{0.99,0.5,0}

\usepackage{cleveref}
\crefname{table}{Table}{Tables}
\crefname{equation}{Eq.}{Eqs.}
\crefname{appendix}{App.}{Apps.}
\crefname{section}{Sec.}{Secs.}
\crefname{figure}{Fig.}{Figs.}

\begin{document}

\vspace{-1cm}

\title{Nuclear Atomic Dark Matter}

\author{Pouya Asadi\,\orcidlink{0000-0003-4712-402X}}
\email{pasadi@ucsc.edu}
\affiliation{Department of Physics and Santa Cruz Institute for Particle Physics, University of California Santa Cruz, Santa Cruz, CA 95064, USA}

\author{Zackaria Chacko\,\orcidlink{0000-0002-7321-8073}}
\email{zchacko@umd.edu}
\affiliation{Maryland Center for Fundamental Physics, Department of Physics, University of Maryland, College Park, MD 20742, USA}

\author{David Curtin\,\orcidlink{0000-0003-0263-6195}}
\email{dcurtin@physics.utoronto.ca}
\affiliation{Department of Physics, University of Toronto, Canada}

\author{Micah Mellors\,\orcidlink{0009-0001-7786-9703}}
\email{m.mellors@mail.utoronto.ca}
\affiliation{Department of Physics, University of Toronto, Canada}

\author{Ethan T.~Neil\,\orcidlink{0000-0002-4915-3951}}
\email{ethan.neil@colorado.edu}
\affiliation{Department of Physics, University of Colorado, Boulder, CO 80309, USA}

\begin{abstract}

The matter coincidence, $\Omega_{\rm DM}\simeq5\Omega_b$, suggests a common origin for visible and dark matter (DM). 
We show that this motivates the study of nuclear atomic DM (naDM), which is composed of 
$\sim5\,\mathrm{GeV}$ mirror-neutron DM and a subdominant population of mirror elements.
We realize naDM in the Mirror Twin Higgs, employing a twin Froggatt--Nielsen mechanism to generate fermion masses and realize spontaneous $\mathbb{Z}_2$ symmetry breaking.
However, the low-energy mirror framework is more general and can emerge from other ultraviolet completions.
naDM gives rise to a novel two-stage dark acoustic oscillation signature, multiple correlated direct-detection signals, and highly exotic twin astrophysics.

\end{abstract}

\maketitle

\textit{\textbf{Introduction} ---} 
The nature of dark matter (DM) remains one of the most pressing questions in physics; see Refs.~\cite{Cirelli:2024ssz,Bozorgnia:2024pwk} for recent reviews. 
While minimal candidates such as weakly interacting massive particles (WIMPs) remain compelling, the matter coincidence problem ($\Omega_{\rm DM}\simeq 5\Omega_b$~\cite{Planck:2018vyg, ParticleDataGroup:2024cfk}) points toward richer dark sectors. 
If visible and dark asymmetries share an ultraviolet (UV) origin, comparable number densities imply DM masses near the hadronic scale and, for a confining dark sector, a confinement scale near that of quantum chromodynamics (QCD).
This motivates models in which a symmetry relates the Standard Model (SM) to the dark sector. 

The Mirror Twin Higgs (MTH) framework~\cite{Chacko:2005pe,Barbieri:2005ri,Chacko:2005vw} addresses the little hierarchy problem through an approximate \Ztwo\ symmetry between the SM and a mirror sector, whose breaking is required by cosmological~\cite{Chacko:2016hvu,Craig:2016lyx,Goldstein:2026iuu} and collider constraints~\cite{Craig:2015pha,Cheng:2015buv,Cheng:2016uqk,Chacko:2019jgi,ParticleDataGroup:2024cfk}. Baryogenesis above the \Ztwo-breaking scale can yield comparable visible and twin asymmetries~\cite{Davoudiasl:2010am,Bell:2011tn}, which, together with related confinement scales, correlate their energy densities. With unbroken twin electromagnetism, charged twin protons ($p_B$) are disfavoured as dominant DM~\cite{Markevitch:2003at,Fan:2013yva,Randall:2008ppe}, while neutral twin neutrons ($n_B$) remain viable, motivating \Ztwo-breaking in the fermion masses that inverts the twin-nucleon spectrum.

A number of mirror-sector models realize such a $n_B$ DM scenario, e.g. Refs.~\cite{Farina:2015uea,Lonsdale:2018xwd,Beauchesne:2020mih,Bodas:2024idn}.
In this work, we consider a concrete realization based on a Froggatt--Nielsen (FN) extension~\cite{Froggatt:1978nt} of the MTH framework~\cite{Barbieri:2017opf}: a modest difference between the two flavon vacuum expectation values (vevs) spontaneously breaks \Ztwo, reshapes the twin-quark spectrum, and makes the $n_B$ the lightest dark baryon and a viable DM candidate, realizing twin neutron DM (TNDM). By marrying a solution to the little hierarchy problem (MTH) with one to the fermion hierarchy problem (FN), this model also addresses the matter coincidence problem.

In general, TNDM can be accompanied by a residual population of stable $p_B$, 
realizing a version of atomic DM (aDM) \cite{Kaplan:2009de}
with nuclear physics that we call 
\textit{nuclear atomic dark matter} (naDM).
During twin Big Bang nucleosynthesis (BBN), the $p_B$ are processed into twin nuclei (from twin deuterium up to twin carbon). 
If the twin electron is relatively heavy, recombination is inefficient, and the twin nuclei may persist as ions until late times. 
If recombination is efficient, twin electron capture will destabilize atomic twin hydrogen, while some heavier elements may remain stable. 
This implies that a twin atomic population, if present, will be composed of only heavier elements.
In this work, we initiate the study of the consequences of naDM as a subdominant DM component. 
These consequences depend primarily on the IR parameters of the twin sector; thus, they 
could be generated by different UV realizations of the mirror symmetry~\cite{Foot:2003jt,Bai:2013xga,Farina:2015uea,Lonsdale:2018xwd,Beauchesne:2020mih,Bodas:2024idn}.

Twin nuclei are electromagnetically coupled to the twin photon bath, while residual nuclear forces couple them to the dominant TNDM. 
This produces characteristic \textit{two-stage dark acoustic oscillations} (DAOs), where in first stage the entire DM population oscillates, followed by an epoch where only the naDM fraction oscillates.
This gives rise to distinctive features in the matter power spectrum that may allow it to be distinguished from more conventional DAOs \cite{Cyr-Racine:2012tfp,Cyr-Racine:2013fsa, Buckley:2014hja,Agrawal:2017rvu,Bansal:2022qbi,Barron:2025dys}.
The naDM and TNDM components also give rise to a host of correlated direct detection signals, as well as distinctive astrophysical and collider signals, some of which are discussed in our
companion paper~\cite{PRD}.

%=======================================================================================================

\medskip

\textit{\textbf{Model} ---} 
We consider an MTH model in which the \Ztwo\ is broken spontaneously by scalars that also act as FN flavons, making the $n_B$ the lightest stable baryon. 
We assume a \Ztwo-symmetric high-temperature baryogenesis and unbroken twin electromagnetism, with some source of late-time asymmetric reheating~\cite{Chacko:2016hvu, Craig:2016lyx} to ensure $\Delta N_\mathrm{eff} \lesssim \mathcal{O}(0.1)$. We emphasize that this construction is not unique, and other models, e.g. Ref.~\cite{Bodas:2024idn}, could give similar IR phenomenology while also addressing the hierarchy problem.

Under the FN model, fermions are charged under a new global $\Uone_X$ which forbids the Yukawa interactions except for the top quark. The effective low energy Yukawas for the lighter quarks arise when $\Uone_X$ is spontaneously broken by the vev of a scalar field $\phi$ called the flavon. Without loss of generality, we take the FN charge of $\phi$ (SM Higgs) to be $X_\phi=1$ ($X_H=0$).
The lighter quark Yukawas can then be built out of higher-dimensional operators, yielding 
\begin{equation}
    Y^{u,d}_{ij} = c^{u,d}_{ij} \epsilon^{|X_{Q_i} - X_{{u,d}_j}|} \ ,
\end{equation}
where $i,j$ are flavour indices, $\epsilon \equiv \langle \phi \rangle/\Lambda_F \ll 1$, $\Lambda_F$ is a heavy mass scale, and $c^{u,d}_{ij}$ are some $\OfOrder{1}$ coefficients. In the near-diagonal limit, the effective quark Yukawas are $y_i \simeq c^{u,d}_{ii}\epsilon^{|X_{Q_i} - X_{{u,d}_i}|}$. 
In the SM sector, this generates the order of magnitude hierarchies for the fermion masses.

If $|X_{Q_1} - X_{{u}_1}| = |X_{Q_1} - X_{{d}_1}| + 1$,
as found in approximately half of the %preferred 
natural charge assignments in Ref.~\cite{Cornella:2023zme}, then $m_u/m_d \simeq \epsilon c_{11}^u/c_{11}^d$. 
Thus, to reproduce the SM masses, there must be a modest hierarchy $c_{11}^d/c_{11}^u\sim 2\epsilon$, which is required for fitting even some of the most natural models to the SM.
Assuming a fully $\mathbb{Z}_2$-symmetric twin extension of the FN setup, spontaneous $\mathbb{Z}_2$ breaking in the flavon vevs of the twin (B) and visible (A) sector with $\epsilon_B \sim 2 \epsilon_A$
will allow the twin up and down quark masses to become equal, 
beyond which the neutron is the lightest stable baryon.
However, the steep 
$\sim \epsilon_{A,B}^{6,7}$
dependence of the up and down masses means the light twin quarks become heavier by two orders of magnitude.
Since the top quark is neutral under $\Uone_X$ to allow $y_t \sim 1$, its mass is barely affected and the twin solution to the little hierarchy problem remains intact. 
The modified twin quark spectrum affects the twin confinement scale via running effects, enabling a heavy enough stable $n_B$ to explain the DM abundance today when we assume equal asymmetric number densities on the two sides.

The low energy phenomenology is largely controlled by a small subset of parameters. The primary quantities of interest are the twin neutron--proton ($n_B$--$p_B$) mass splitting $\Delta m_{np,B}$, the twin electron mass $m_{e,B}$, the twin Fermi constant $G_{F,B}$, and the temperature ratio between the twin and SM sectors $\xi \equiv T_B/T_A$. 
We focus on the case where the average twin nucleon mass is $m_{N,B}\sim 5\gev$, so as to most directly solve the coincidence problem. For twin up/down quarks with masses $\sim 100\mev$ as motivated by the FN construction, the twin pion mass is $m_{\pi,B}\sim 2- 2.5\gev$. The usual moderate tuning of the twin Higgs vev $f/v\sim 3-9$ \cite{Burdman:2014zta}, where $v$ is the physical Higgs vev, implies $10^{-2}\lesssim G_{F,B}/G_{F,A} \lesssim 10^{-1}$. In our benchmark, we assume asymmetric reheating gives rise to $\xi= 0.2$, which corresponds to $\Delta N_{\rm eff}\simeq 0.01$.

In obtaining the naDM abundance from these parameters, the dominant uncertainty is the twin-QCD dependence of nuclear binding energies, particularly $B_{D,B}$ (deuterium) and $B_{He,B}$ (helium) (see Ref.~\cite{PRD} for detailed discussions). To estimate these energies quantitatively, we make use of lattice results at heavier pion masses which suggest values above their SM counterparts~\cite{NPLQCD:2012mex,Yamazaki:2012hi,NPLQCD:2013bqy,Yamazaki:2015asa,Berkowitz:2015eaa,NPLQCD:2020lxg}, with $50\mev\lesssim B_{D,B}\lesssim180\mev$~\cite{Savage:2015eya} and $100\mev\lesssim B_{He,B}\lesssim350\mev$~\cite{Kirscher:2015yda}.\footnote{Reflecting the highly challenging systematics of these calculations, more recent lattice results~\cite{BaryonScattering:2025ziz,Detmold:2026vjx,Detmold:2026shh,Wang:2026kab,Chakraborty:2026zeq,Zhang:2026cco,BaryonScattering:2026voj} suggest that di-nucleon states may instead be unbound at heavier pion masses, restricting formation of heavier twin elements; the two-scale DAOs and $n_B$ Higgs portal direct detection signals remain in either case.  We analyze both possibilities in Ref.~\cite{PRD}.}  These values suggest primordial synthesis beyond twin helium, motivating the qualitative discussion of twin-element signatures below.

\medskip 
%=======================================================================================================

\textit{\textbf{Nuclear Atomic Dark Matter} ---} 
In addition to the $n_B$, $p_B$ or twin atoms can constitute a fraction of the DM if they are stable. 
Note that depending on the twin baryon mass splitting and the twin electron mass, the $p_B$ may be stable or unstable. If the twin electron is charged under the same FN as the quarks, it is likely heavy $m_{e_B} \sim 0.1~\mathrm{GeV}$~\cite{Cornella:2024jaw}, 
making the $p_B$ stable for all but the largest mass splittings. However, as in the SM, the proton can be stabilized when bound into a nucleus, so a relic population can also exist in the unstable case. 

To calculate the relic abundance of twin atoms, we estimate the twin-BBN relic abundance. 
For nucleosynthesis to proceed, the twin deuteron must be bound and stable, requiring
\begin{equation}\label{eq:stability_cond_D}
    B_{D,B}> -\Delta m_{np,B} - m_{e,B} \, ,
\end{equation}
where $\Delta m_{np,B} \equiv m_{n,B}-m_{p,B}<0$ is the $n_B$--$p_B$ mass splitting.
To characterize the low energy population of DM, we must consider $n_B$--$p_B$ freezeout. 
This process will end when photodissociation of deuterons becomes too slow to prevent nucleosynthesis, and the relic protons are fused into deuteron nuclei. The equilibrium abundance of deuterons is proportional to $\exp(B_{D,B}/T_B)$, and we find, to a good approximation, that 
\begin{equation}
    \frac{T_{D,B}}{B_{D,B}}
    \simeq
    \frac{T_{D,A}}{B_{D,A}}\, ,
    \label{eq:T_bottle_from_SM}
\end{equation}
where $T_D$ is the temperature at which the deuterium bottleneck resolves and a substantial population of nuclei can form. The binding energy range implied by the lattice results \cite{Savage:2015eya} suggests $T_{D,B}\sim1-5\mev$. We make the approximation that at $T_{D,B}$, all remaining relic protons are instantly bound into nuclei.

We follow the treatment of Ref. \cite{Mukhanov:2003xs}, appropriately rescaled for the twin sector, to determine the $n_B$--$p_B$ freezeout. 
If the twin electron is light, twin freezeout proceeds in the same manner as SM freezeout, and the proton population levels off when the weak nucleon-conversion rates drop below the Hubble rate.
If the twin electron is heavy, then nucleon-conversion reactions become inefficient as the population of electrons and positrons drops. 
Freezeout occurs when either the electrons deplete or the weak rate becomes slower than Hubble (see App.~\ref{app:details} for more details), whichever occurs first. The relic $p_B$ fraction depends on the twin neutron-proton conversion freeze-out temperature $T_\mathrm{dec}$ as $X_{p,B} \equiv n_{p,B}/n_{b,B}\simeq \exp(- |\Delta m_{np,B}|/T_{dec})$.

\begin{figure}
    \centering
    \includegraphics[width=0.49\textwidth]{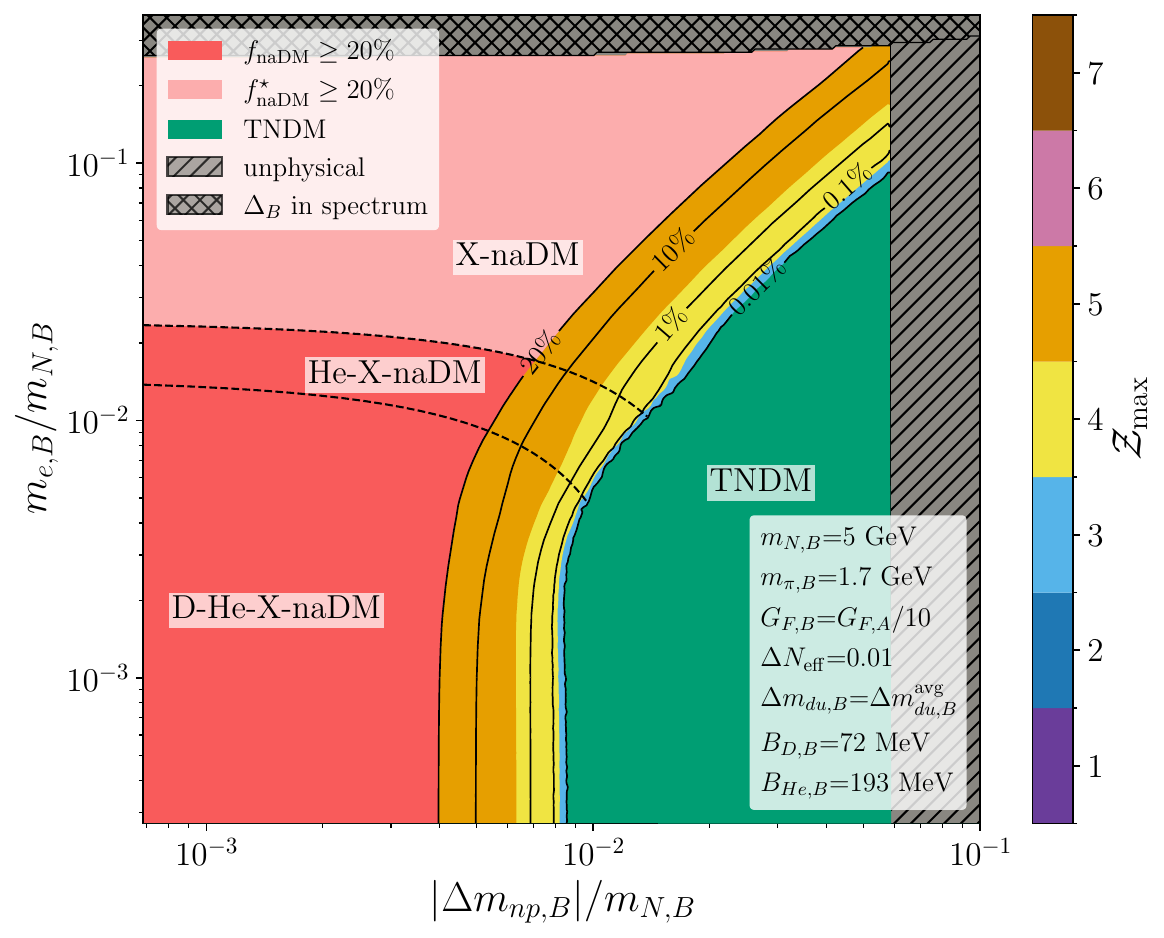}
    \caption{
    Twin nuclei DM fraction $\fnaDM$ determined using rescaled SM freezeout expressions from Ref. \cite{Mukhanov:2003xs}, given in the labelled contours. The overlaid dashed lines indicate regions where twin heliums (and deuteriums) are stable to decay via electron capture, indicated as (D-)He-X-naDM; X-naDM denotes the region where both are unstable. 
    \textcolor{N_SIDM}{Green} indicates the region where the population is effectively pure TNDM, either because the twin deuteron is unstable to decay via \cref{eq:stability_cond_D} or because the $p_B$ fraction at $T_{D,B}$ is small, $\fnaDM< 10^{-4}$. The \textcolor{red}{red} region is (naively) excluded by self-interaction considerations with $\fnaDM\gtrsim 20\%$.
    In the \textcolor{red!50}{pink} region indicated by $\fnaDM^\star$, atoms may decay after recombination, reducing or eliminating the late-time naDM fraction. In principle, relic ions could be produced in all regions except pure TNDM.
    The remaining colours indicate the expected value of $\Znum^{\max }$. The grey region is either unphysical or out of scope of this analysis (as we neglect twin $\Delta$ baryons). 
    The parameters at this benchmark are defined in the legend; the exact location of the contours will change as these are varied (see Ref.~\cite{PRD}) but the qualitative picture remains the same. 
    }
    \label{fig:freezeout}
\end{figure}

Bullet Cluster observations constrain the \emph{late-time} long-range-interacting naDM fraction to $\fnaDM\lesssim10$--$20\%$~\cite{Markevitch:2003at,Fan:2013yva}.
We therefore focus on parameters where $p_B$ is sufficiently depleted before nucleosynthesis. An appreciable relic fraction, $\fnaDM\sim10\%$, occurs for
\begin{equation}\label{eq:freezeout_mass_splitting}
\begin{aligned}
    \left|\Delta m_{np,B}\right|_{L}&\sim 
    50\mev 
    \left(\frac{G_{F,A}/10}{G_{F,B}}\right)^{2/3}
    \left(\frac{0.2}{\xi}\right)^{2/3},
    \\
    \left|\Delta m_{np,B}\right|_{H}&\sim m_{e,B}/2,
\end{aligned}
\end{equation}
where $L$ and $H$ denote the light- and heavy-electron regimes. The benchmark freezeout fractions in \cref{fig:freezeout} exhibit these two scalings in the $(\Delta m_{np,B},m_{e,B})$ plane.

Lacking a reliable rescaled twin-BBN reaction network, we estimate heavy-nucleus formation using the geometric rates of Refs.~\cite{Krnjaic:2014xza,Hardy:2014mqa}. 
Evaluating the Coulomb barrier of fusing nuclei at $T_{D, B}$ yields an upper bound on the atomic number $\Znum \leq\Znum_{\max}$ of produced twin nuclei.
Taking twin baryons (protons) to constitute $100\%$ ($10\%$) of DM gives
\begin{equation}\label{eq: heavy nuclei Zmax}
    \Znum_{\max}\sim 5
    \left(\frac{B_{D,B}}{100\ \textrm{MeV}}\right)^{1/5}
    \left(\frac{\alpha_A}{\alpha_B}\right)^{2/5}
    \left(\frac{5\ \textrm{GeV}}{m_{N,B}}\right)^{1/5},
\end{equation}
up to logarithmic corrections, see App.~\ref{app:details}. As the proton freezeout abundance decreases, the reduced reactant density lowers $\Znum_{\max}$, as shown in \cref{fig:freezeout} and \cref{eq:fusion_rate}. This estimate is only an upper bound, since unbound intermediate states may halt fusion.

The flipped twin neutron-proton mass ordering may make some of the resulting atoms unstable to twin electron capture,
${}^{A}_{Z}X_B+e_B^-\to{}^{A}_{Z-1}Y_B+\nu_B$.
Twin recombination, if it is efficient, can therefore 
transmute some or all of the atomic subcomponent to lower-$\Znum$ isobars or even entirely into TNDM.
If twin electrons are heavy, twin recombination will be inefficient~\cite{Kaplan:2009de} and a sizeable population of relic twin ions (including $p_B^+$) may persist to late times.
In \cref{fig:freezeout}, we indicate the regions where only twin D atoms and both twin D and He atoms are unstable. Twin H atoms are always unstable. The stability of heavier atoms in the X-naDM regime has to be established by future studies.

We can therefore imagine two scenarios where the \emph{primordial} $\fnaDM$ fraction could be larger than 20\%: (1) If the heavy elements are all unstable to electron capture, they do not survive to late times. (2) If the heavy elements are stable, $\fnaDM \gtrsim 20\%$ implies $\Znum_\mathrm{max} \gtrsim 6$, possibly resulting in the formation of 
\textit{primordial twin dust}.
In the former (latter) case, large or dominant early-time (late-time) fractions of naDM might therefore evade the nominal self-interaction bound.

\medskip 

%=======================================================================================================
\textit{\textbf{Signals} --- }
At early times (before recombination), the massless twin photon is tightly coupled to the charged twin constituents---free twin electrons and protons/twin nuclei---so that radiation pressure supports acoustic waves in the charged naDM component. 
Unique to the naDM scenario, prior to twin neutron-proton decoupling at $T_{np,\mathrm{dec}}$, twin strong interactions  transfer momentum between the charged nuclei and the electromagnetically neutral $n_B$, dragging the TNDM component with the photon-coupled plasma. 
This realizes the novel two-step DAO signature, with all the DM oscillating for $T > T_{np,\mathrm{dec}}$, and only the charged component $f_{\rm naDM}\lesssim0.1$ oscillating thereafter until recombination (if efficient) or Compton decoupling.
This has no analogue in a single-component collisionless relic, minimal aDM, or previous works on DAO \cite{Cyr-Racine:2012tfp,Cyr-Racine:2013fsa,Buckley:2014hja,Agrawal:2017rvu,Bansal:2022qbi,Barron:2025dys}.

For twin hydrogen, the electronic binding energy is $B^e_{H,B}\sim\alpha_B^2 m_{e,B}/2$. 
We will take as our default assumption that recombination is efficient at depleting the charged population, (as is the case in most of our parameter space); alternative scenarios are considered in our companion paper~\cite{PRD}.
Defining $T_{\rm rec}$ as the visible-sector temperature at which the free twin-electron fraction falls below $10\%$, the Saha equation gives 
\begin{equation}
T_{\rm rec}\sim
  1.5 \ev \left(  \frac{\alpha_B^2m_{e,B}}{\alpha_A^2 m_{e,A}}\right)\left(\frac{0.2}{\xi}\right)\,.
\label{eq:DAO_recombination}
\end{equation}
If deuterium, helium, or a heavier element dominates naDM, the changes in the Saha prefactors are only $\mathcal{O}(1)$, while the larger ionization energy raises $T_{\rm rec}$. 
Note that in general, $T_{\rm rec}\ll ~1$~keV.
The strong interactions keeping $n_B$ and $p_B$ in kinetic equilibrium decouple at higher temperatures $T_{np,\mathrm{dec}}\sim \kev \gg T_\mathrm{rec}$, meaning the neutron--ion decoupling will generally determine when the dominant TNDM component ceases to oscillate, see App.~\ref{app:details} for details.
The corresponding present-day wave number at $T_{np,\mathrm{dec}}$ is
\begin{equation}
\begin{aligned}
k_{np,{\rm dec}}
&\sim 50\, h\,{\rm Mpc}^{-1}
\left(\frac{m_{N,B}}{5\,{\rm GeV}}\right)
\left(\frac{0.05}{X_{p,B}}\right)^{2/3}\\[-2pt]
&\qquad\times
\left(\frac{10\,{\rm GeV}^{-1}}{L_B}\right)^{4/3}
\left(\frac{0.2}{\xi}\right)^{1/3}.
\end{aligned}
\label{eq:DAO_scale}
\end{equation}
Here $X_{p,B}=0.05$ corresponds approximately to $f_{\rm naDM}=0.1$ if nuclei heavier than hydrogen form. The benchmark twin nucleon interaction length $L_B=10\,{\rm GeV}^{-1}$ is a fm-scale interaction length, as expected for a strong force. 
The result depends only weakly on $\xi$ but strongly on $L_B$, motivating sharper lattice calculations of $L_B$. 
The exact scales involved will also depend on the atomic composition, as the number densities and interaction lengths will differ. 

\begin{figure}
    \centering
    \includegraphics[width=1\linewidth]{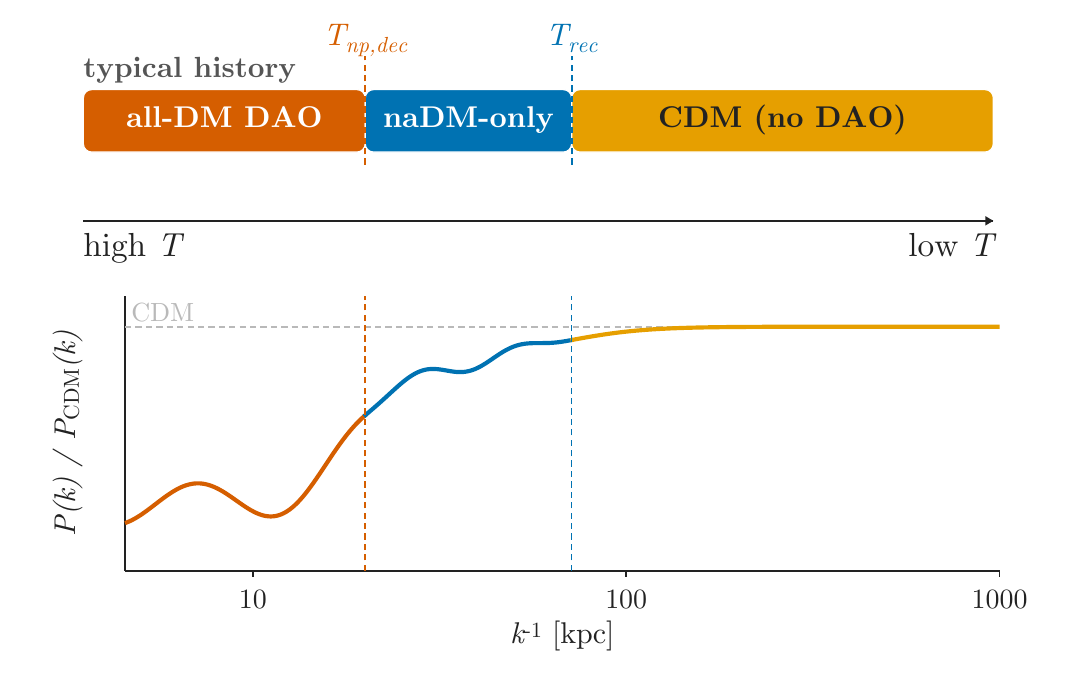}
    \caption{
    Schematic DAO evolution and its impact on the small-scale power spectrum. \textbf{Top:} Thermal history the typical ordering of recombination and kinetic decoupling. \textbf{Bottom:} Corresponding schematic matter-power spectrum, with colors matching the epochs in the top panel. The two-stage DAO in our model gives rise to a unique imprint on the matter power spectrum, potentially detectable in future surveys. 
    }
    \label{fig:dao}
\end{figure}

\cref{fig:dao} gives a schematic of the resulting DAOs.
All of DM undergoes DAOs until $n_B$--$p_B$ decoupling at $T_{np,{\rm dec}}$, after which the neutrons behave as (approximately) collisionless DM while the residual naDM continues to oscillate until recombination at $T_{\rm rec}$. 
Since the DAO signal arises from the dynamics before recombination, this signal exists for the entire viable naDM region from \cref{fig:freezeout}, regardless of whether naDM atoms exist today or decayed to TNDM after recombination.

This strong-interaction amplification makes naDM qualitatively different from conventional aDM. 
If only naDM oscillates, current ultraviolet luminosity function measurements approximately require $k_{\rm DAO}\gtrsim 20\,h\,{\rm Mpc}^{-1}$ or $f_{\rm naDM}\lesssim0.07$ \cite{Barron:2025dys}.
When strong interactions drag the neutrons, the corresponding all-DM requirement strengthens to $k_{\rm DAO}\gtrsim 50\,h\,{\rm Mpc}^{-1}$~\cite{Yuan:2026zlr}. 
This approaches galactic scales, which are beyond most standard cosmological probes, but could be explored by, for example, the Milky Way sub-halo mass function and strong lensing~\cite{Enzi:2020ieg,Nadler:2021dft}.
Our results motivate dedicated analyses of DAOs for these observables, alongside new possible search strategies, e.g. in 21cm astronomy.

\begin{figure}
    \centering
    \includegraphics[width=\linewidth]{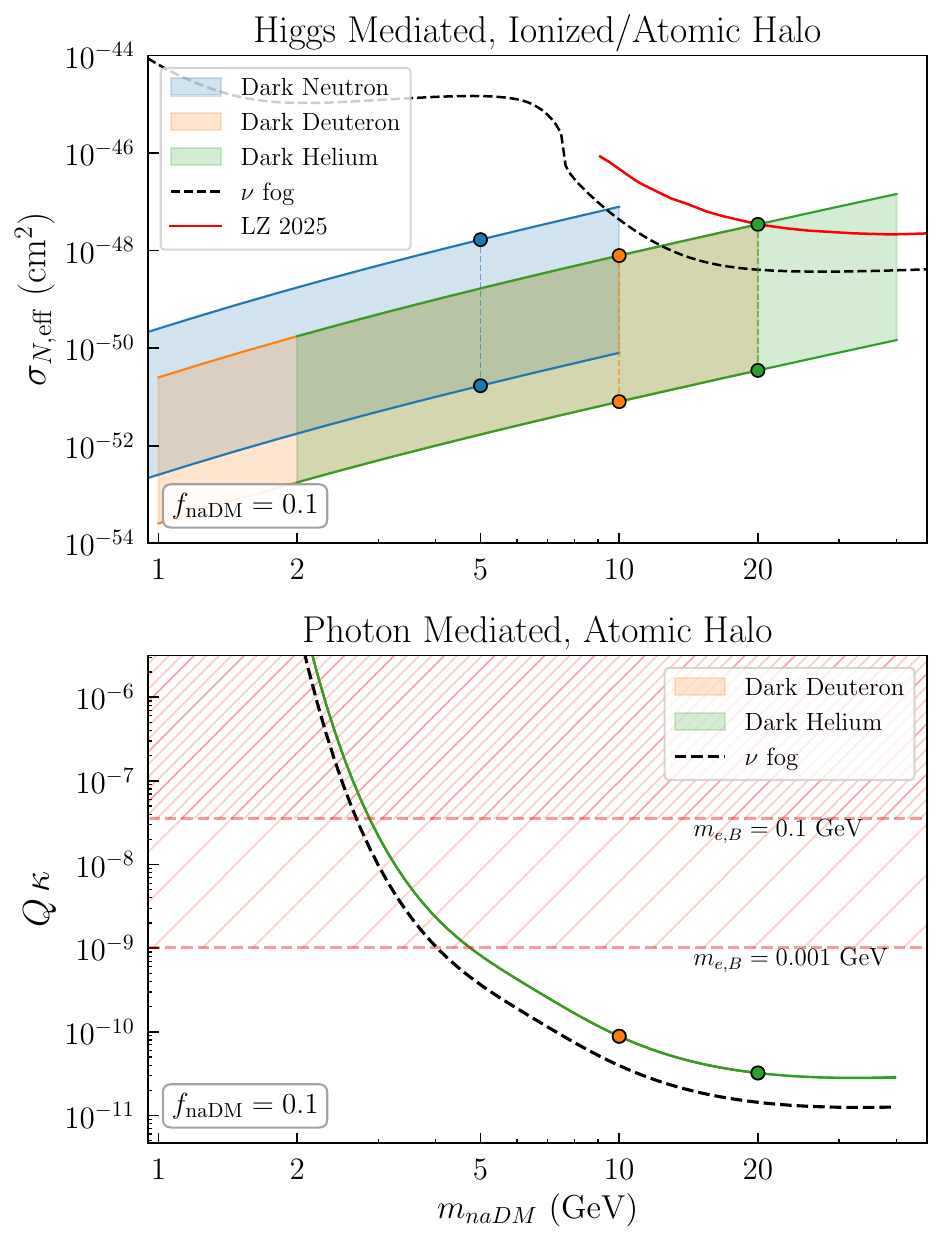}
    \caption{
    \textbf{Top}:
    Predicted Higgs-mediated direct detection cross section for TNDM and some naDM candidates. The effective cross section $\sigma_{N,\mathrm{eff}}$ is weighted by the relative abundance of the DM component. The shaded bands arise from scanning a range of values for $G_{F,B}$ and $0.14\lesssim m_{\pi,B}/m_{N,B}\lesssim 0.56$, where varying $m_{\pi,B}$ modifies the underlying quark masses which set the TNDM-Higgs coupling. Constraints shown are from LUX-ZEPLIN (LZ) \cite{LZ:2024zvo} and the neutrino fog is taken from Ref.~\cite{OHare:2021utq} for WIMP DM in a SHM velocity distribution.
    \textbf{Bottom}:
    LZ constraints on $Q\kappa$ from photon-mediated scattering of atomic naDM, with $Q$ the nuclear charge, requiring less than three signal events.
    \textcolor{red}{Red} dashed lines show the cosmological bounds on $\kappa$~\cite{Vogel:2013raa} for $m_{e,B} = 0.001~\mathrm{GeV}$ and $m_{e,B} = 0.1~\mathrm{GeV}$, assuming $Q=1$.
    The naDM candidates are fixed at 10\% of the total DM population for both plots.
    Dots indicate the coincidence solution points, $m_{N,B}\simeq5\gev.$
    }
    \label{fig:direct_detection}
\end{figure}

In some parts of our parameter space, the $p_B$ is completely depleted, leaving only TNDM in the bath.
Despite no DAO signals, this part of the parameter space still gives rise to Higgs-mediated direct detection signals. 
The predicted direct detection cross section is given in \cref{fig:direct_detection}. The neutron signal itself often lies within the solar neutrino fog, making it a particularly well-motivated target for directional detectors, e.g. see Refs.~\cite{Grothaus:2014hja, OHare:2015utx, Vahsen:2020pzb, Sassi:2021umf}. We note that at the point motivated by the MTH--FN construction, twin quarks are heavy and the TNDM-SM Higgs coupling is at the upper end of the predicted range for a given $G_{F,B}$. 
For twin nuclei, coherent scattering enhances the rate with nucleon number, allowing some deuterium- and helium-like candidates to lie above the neutrino fog even after rescaling by their subdominant abundance. The direct detection prospects for heavier twin nuclei are even more promising.

A small kinetic mixing $\kappa F_A^{\mu \nu} F_{B, \mu \nu}/2$~\cite{Holdom:1985ag, Gherghetta:2019coi} between the SM and twin photons gives naDM states an induced \textit{nanocharge}. 
Current constraints are shown in~\cref{fig:direct_detection}. 
For twin electrons $m_{e,B}\lesssim 1$ GeV, the Bohr radius of the twin atom is large enough that the atoms are collisionally trapped within the halo, so the high-velocity tail extends beyond the galactic escape velocity rather than being cut off as in the standard halo model (SHM)~\cite{Lewin:1995rx, Baxter:2021pqo,DelNobile:2021wmp}. This high energy tail drives the signal at low naDM masses. 
We assume a Maxwell-Boltzmann distribution $f(v)\sim \exp(-v^2/v_0^2)$ with $v_0= 220$ km/s and reinstate the galactic escape velocity cutoff $v_{\rm esc}=544$ km/s for the case of heavy electrons $m_{e,B}\gtrsim 100\mev$, where the Bohr radius is small and scattering is suppressed.

For lighter twin electrons, the naDM component may cool efficiently enough to form a dark disk~\cite{Fan:2013yva,Ghalsasi:2017jna, Chang:2018bgx, Huo:2019yhk, Shen:2021frv, Shen:2022opd, Roy:2023zar, Roy:2024bcu}. We consider this and the possibility of a fully ionized halo in our companion paper~\cite{PRD}. Electron recoil signals can be substantial for ionized twin electrons. In the presence of multiple different naDM species we can get a host of different signals at direct detection searches, with masses in intervals of $\sim 5\gev$. 

The model has several other complementary signals. 
Like all aDM scenarios, naDM predicts mirror stars~\cite{
Blinnikov:1982eh, Khlopov:1989fj, Foot:2000vy, Curtin:2019ngc,Curtin:2019lhm,Armstrong:2023cis,Cabral:2026tjb}, which may be identifiable in current astrophysical data from surveys such as \textit{Gaia} \cite{Howe:2021neq,Gaia:2016},  but the twin nuclear physics could prolong their life or give them unique properties.
The SM Higgs decay branching fraction to twin fermions depends on the full spectrum of twin fermion masses, but we show in our companion paper that HL-LHC in conjunction with FCC-ee \cite{ParticleDataGroup:2024cfk,deBlas:2019rxi} will be able to place powerful constraints on MTH-FN twin sectors.

%================================================================================================================================

\medskip

\textit{\textbf{Conclusion} ---} 
In mirror solutions to the hierarchy problem, incorporating a solution to the flavour problem can naturally invert the twin neutron--proton mass hierarchy. The twin neutron then becomes a viable DM candidate, simultaneously addressing the matter coincidence problem. 
A generic consequence of these scenarios is a subdominant population of twin nuclei in the dark sector. In this work, we highlighted a particular UV completion, in which the MTH solution to the little hierarchy problem is combined with the FN explanation of the fermion mass hierarchy, making the twin neutron a viable $\sim 5\gev$ DM candidate. Supplemented by \Ztwo-symmetric asymmetry generation and the usual asymmetric reheating, this construction correlates the visible and dark matter abundances. The benchmark parameters considered here are motivated by this UV completion. Our companion work~\cite{PRD} maps out the general IR parameter space, making our results applicable  to other models which can realize naDM~\cite{Foot:2003jt,Bai:2013xga,Farina:2015uea,Lonsdale:2018xwd,Beauchesne:2020mih,Bodas:2024idn}. 

We map the low-energy parameter space in which $n_B$--$p_B$ freezeout can yield a viable naDM subfraction and find a preferred band in the $m_{e,B}-\Delta m_{np,B}$ plane. Dominant twin-neutron DM can coexist with subdominant naDM, whose residual nuclear interactions can drive a unique two-stage DAO to produce observable small-scale structure suppression and oscillations. Direct detection, invisible Higgs decays, and further astrophysical signatures (e.g. crystallization of twin dust or formation of mirror stars) provide complementary probes. Some aspects of the broader phenomenology and a supersymmetric UV completion are studied in the accompanying work~\cite{PRD}.

Our framework solves the matter coincidence problem as a consequence of combining solutions to the little hierarchy and flavour problems. 
This serendipity, alongside its unusually rich set of predicted signals, motivates future work on its theoretical underpinnings and detailed signatures at all scales.

\begin{acknowledgments}
We thank Wolfgang Altmannshofer, Stefania Gori, Michael Luke, Jared Barron, and Nick Rodd for helpful discussions. 
The work of PA is supported in part by the U.S. Department of Energy grant number DE-SC0010107.
ZC is supported in part by the National Science Foundation
under Grant Number PHY-2514660 and by the Maryland Center for Fundamental Physics.
The work of DC is in part supported by Discovery Grants from the Natural Sciences and Engineering Research Council of Canada, the Canada Research Chair program, the Ontario Early Researcher Award, and the University of Toronto McLean Award.
The work of MM is also supported in part by the NSERC Postgraduate Scholarship-Doctoral.
The work of ETN is supported in part by the U.S. Department of Energy under grant number DE-SC0010005.
In our analysis and preparation of this manuscript, Claude and ChatGPT were used to assist with literature searches, resolve stubborn \LaTeX{} errors, refine the language, and iterative discussions and clarifications of selected physics questions. 
The authors take full responsibility for the content of the manuscript.
\end{acknowledgments}

\appendix

%================================================================================================================================
\section{Details of Twin Sector Calculations}\label{app:details}

Here we briefly derive the relations reported in the main text. 

%================================================================================================================================
\textit{\textbf{Twin Neutron-Proton Freezeout} ---} In our detailed numerics, we follow Ref.~\cite{Mukhanov:2003xs} for the interaction rates. Here we show the estimates for the decoupling temperature in the two regimes discussed in the main text. 

Twin weak processes such as $p_B + \bar{\nu}_B \leftrightarrow n_B + e_B^+$ maintain the twin proton abundances at their equilibrium values, $X_{p,B}\simeq \exp{\left(-|\Delta m_{np,B}|/T_B\right)}$ so long as the rates are faster than the Hubble rate. Below asymmetric reheating, the expansion rate is set by the SM bath, $H\sim T_A^2/\Mpl$. If the twin electron is light, the weak rates responsible for converting twin baryons are expected to scale as $\Gamma\sim G_{F,B}^2 T_{B}^5$, in direct analogy with the SM. Setting $\Gamma=H$, we find the twin sector decoupling occurs at an SM photon temperature of 
\begin{equation}\label{eq:Tfreezeout1}
    T_{B,\mathrm{dec},1}
    \sim
    \left(\Mpl G_{F,B}^2\xi^2
    \right)^{-1/3}\, .
\end{equation}

If the twin electron is heavy, then twin nucleon-conversion reactions become inefficient once the $e^\pm_B$ population becomes Boltzmann suppressed, at 
\begin{equation}\label{eq:Tfreezeout2}
    T_{B,\mathrm{dec},2}
    \sim
    \frac{m_{e,B}}{x_e}\, ,
    \qquad
    x_e\sim \mathcal{O}(5\text{--}10)\, ,
\end{equation}
Freezeout occurs at whichever temperature is higher $T_{dec}= \max(T_{B,dec,1}, T_{B,dec,2})$. Setting the abundance $X_{p,B}\simeq \exp(-|\Delta m_{np,B}|/T_{dec})$ equal to 10\% and solving for $\Delta m_{np,B}$ for each decoupling temperature yields the relations in~\cref{eq:freezeout_mass_splitting}. Note that if $T_{D,B}$ exceeds both decoupling temperatures, the proton population relevant for BBN is set by the deuterium temperature, though at our benchmark point this is not the case.

%================================================================================================================================
\medskip
\textit{\textbf{Twin Nucleosynthesis} ---}  
Once the twin baryons are nonrelativistic, their scattering can be parameterized by the twin nucleon scattering length $a_B$ ~\cite{Bethe:1949yr, Blatt:1949zz,Chu:2019awd}. 
The cross section scales as $\sigma_s\simeq\pi a_B^2$ for $a_B\ll(m_{N,B}v)^{-1}$ and approaches $\sigma_s\simeq\pi/(m_{N,B}v)^2$ in the opposite limit. We capture both regimes by defining
\begin{equation}\label{eq:interaction_length}
L_B^2\equiv
\min\!\left[a_B^2,(m_{N,B}\xi T_A)^{-1}\right].
\end{equation}
A more careful determination of this interaction length scale requires convergence on lattice calculations which are not yet settled for these multi-nucleon parameters. For heavier nuclei, the length scale should be rescaled according to the nuclear radius $r_\Anum \propto \Anum^{1/3}$, with $\Anum$ the atomic mass number. 

The freezeout temperature of twin nuclear fusion rates is largely set by the temperature at which the reactants can no longer overcome the Coulomb barrier, determined by the fusion rate~\cite{Krnjaic:2014xza, Hardy:2014mqa}
\begin{equation}\label{eq:fusion_rate}
\begin{aligned}
        \Gamma_{fusion}&\simeq n_\Anum \langle\sigma_\Anum v_\Anum \rangle\, e^{-\sqrt{E_G/E}},
        \\
        &\simeq 2 X_{p,B}n_{b,B}\langle \sigma_1 v_1 \rangle \Anum_i^{5/6} e^{-\sqrt{E_G/E}},
\end{aligned}
\end{equation}
where $n_{\Anum}$ is the number density, $\sigma_{\Anum}$ is the cross section, and $v_{\Anum}$ is the velocity of incoming nuclei with atomic mass number (proton number) $\Anum$ ($\Znum$). Here we have assumed that the interaction length can be rescaled geometrically from the nucleon interaction length $L_B$, the velocity is rescaled according to thermalization of the species in equilibrium, and that a single reactant species $\Anum_1=\Anum_2=\Anum_i\simeq 2\Znum_i$ dominates the abundance at the temperature of interest. $E\simeq 3T_B/2$ is the typical energy of the incoming particles and $E_G$ is the Gamow energy \cite{Maoz:2016textbook}
\begin{equation}
    \frac{E_G}{E} \simeq \frac{4 \pi^2\Znum_i^5 \alpha_{EM,B}^2 m_{N,B}}{3 \,T_{B}},
\end{equation}
with $\Znum_i$ the proton number of the initial state reactants. The late times number density is set by requiring that the twin baryons comprise all of DM
\begin{equation}\label{eq:twin_baryon_number_density}
    n_{b,B}= n_{b,A}= \eta\frac{2\zeta(3) T_A^3}{\pi^2} 
\frac{\Omega_{\rm DM}}{\Omega_b}
\frac{m_{N,A}}{m_{N,B}},
\end{equation}
where $\eta\sim 10^{-10}$  is the measured SM baryon-to-photon ratio~\cite{ParticleDataGroup:2024cfk}. The last two fractions are unity at the benchmark coincidence solution point. 

Solving $\Gamma_{fusion}\simeq H$, with $H$ again dominated by the SM dof, yields the approximate freezeout temperature
\begin{equation}\label{eq: fusion freezeout temp approx}
        T_{FO} \sim    \frac{3 \times 10^{-8}\ \textrm{GeV}\, \Anum_f^5}{\left(1 - 0.07 \ln \Anum_f - \mathfrak{L} \right)^{2}}
    \left(\frac{\alpha_B}{\alpha_A}\right)^2
    \left(\frac{m_{N,B}}{5\ \textrm{GeV}}\right)
    ,
\end{equation}
where we have used $\Anum_f=2\Anum_i$ and $\Anum_f=2\Znum_f$. The logarithmic corrections $ \mathfrak{L}$ are 
\begin{widetext}
\begin{equation}
     \mathfrak{L} \simeq 0.09
    \ln\left(
    \left(\frac{g_{\star,S}}{10.75}\right)^{1/2}
    \left(\frac{m_{N,B}}{5 \gev}\right)
    \left(\frac{\xi}{0.2}\right)
    \left(\frac{0.1}{X_p}\right)
    \left(\frac{1 \gev^{-1}}{L_B}\right)^{2}
    \left(\frac{\mev}{T_{FO}}\right)^{3/2}
    \right).
\end{equation}
\end{widetext}
Increasing $\mathfrak{L}$ increases the freezeout temperature, disfavouring the fusion reaction.
The requirement $T_{FO}<T_D$, with $T_D$ from~\cref{eq:T_bottle_from_SM}, yields the approximate value for $\Znum_{\max}$ reported in~\cref{eq: heavy nuclei Zmax}. The full numerical solution is used for the estimates in~\cref{fig:freezeout}. In this plot we also use the more careful criteria $\Znum= \lfloor\Anum/2\rfloor$ and reactant atomic mass numbers $\Anum_1 = \lfloor \Anum_f/2 \rfloor$, $\Anum_2 = \Anum_f -  \Anum_1$. The choice of floor identifies neutron-rich isotopes. We set the length scale $L_B$ to the nuclear radius $1.2$ fm and rescale with $\Lambda_{\mathrm{QCD},A}/\Lambda_{\mathrm{QCD},B}$ on dimensional grounds. Details of the determination of $\LQCD$ can be found in the companion paper~\cite{PRD}.

We comment finally on the fact that, if binding energies in the twin sector are substantially larger than their SM counterparts, then the twin di-neutron state (marginally unbound in the SM) may be bound and stable and can change some of the signals significantly. For instance, they have larger couplings to the Higgs, enhancing the direct detection cross section---see Ref.~\cite{PRD} for more details on this possibility.

%================================================================================================================================
\medskip
\textit{\textbf{Dark Acoustic Oscillations} ---}
The twin weak interactions decouple above an MeV and are irrelevant for DAOs, whereas the strong interactions remain efficient much later.
Using $v^2\sim T_B/m_{N,B}$ and the late-time twin-baryon density~\cref{eq:twin_baryon_number_density}, the momentum-transfer rate is
\begin{equation}
\begin{aligned}
\Gamma_{s,B}
&=n_{p,B}\langle\sigma v\rangle\\
&\sim \left(X_{p,B}\eta T_A^3
\frac{\Omega_{\rm DM}}{\Omega_b}
\frac{m_{N,A}}{m_{N,B}}\right)
\left( L_B^2
\sqrt{\frac{\xi T_A}{m_{N,B}}}\right),
 \end{aligned}
\label{eq:DAO_rate}
\end{equation}
with $L^2_B$ defined in Eq.~\eqref{eq:interaction_length}. Equating~\cref{eq:DAO_rate} to the Hubble rate gives
\begin{equation}
\begin{aligned}
T_{np,{\rm dec}}
&\simeq 0.6\,{\rm keV}
\left(\frac{m_{N,B}}{5\,{\rm GeV}}\right)
\left(\frac{0.05}{X_{p,B}}\right)^{2/3}\\[-2pt]
&\qquad\times
\left(\frac{10\,{\rm GeV}^{-1}}{L_B}\right)^{4/3}
\left(\frac{0.2}{\xi}\right)^{1/3}.
\end{aligned}
\label{eq:DAO_estimate_T}
\end{equation}
For a DAO length scale $r_{\rm DAO}^{-1}\simeq \sqrt{3}H(T_{dec})$, $k_{\rm DAO}\equiv \pi/r_{\rm DAO}$. The present day scale corresponding to a particular temperature is therefore
\begin{equation}
    k_{\rm DAO,0}\simeq \sqrt{3}\pi H(T_{dec}) \frac{T_{0}}{T_{dec}},
\end{equation}
which yields~\cref{eq:DAO_scale}, where we again assume that at late times Hubble is dominated by SM dof. 

%================================================================================================================================
\medskip
\textit{\textbf{Higgs-Nucleon Couplings} ---} The per-nucleon cross section for Higgs-mediated nuclear recoil is
\begin{equation}\label{eq: Higgs cross section}
   \sigma_N \simeq \frac{\mu_N^2  }{\pi m_{h,A}^4  }
        c_{N,B}^2 
        c_{N,A}^2,
\end{equation}
where $\mu_N$ is the reduced mass of the target species and the TNDM. The couplings $c_{N,A/B}$ of nucleons to the SM Higgs are related to the quark couplings via
\begin{equation}
    \begin{aligned}
        c_{N,A} &\equiv \frac{m_{N,A}}{v}\sum_q  f_{q,A}^{(N)}, 
        \\
        c_{N,B} & \equiv \left(\frac{m_{N,B}}{f} \right) \left(\frac{v}{f}\right) \sum_q  f_{q,B}^{(N)}(m_{\pi,B}) 
        \label{eq:cN_Higgs_portal}
    \end{aligned}
\end{equation}
The nucleon scalar form factor $f_{q,X}^{(N)}$ encodes the quark mass contributions to the nucleon mass and can be extracted from the lattice for the SM \cite{ FlavourLatticeAveragingGroupFLAG:2024oxs, Alexandrou:2014sha,Gupta:2021ahb,Freeman:2012ry} as well as at heavier than physical pion masses \cite{DeGrand:2019vbx}. The SM nucleon effective Yukawa is then $c_{N,A} \simeq 1.2 \times 10^{-3}.$ The twin neutron coupling to the SM Higgs is suppressed by $v/f$, but is partially compensated by having heavier quarks and thus a stronger coupling to the twin Higgs, 
\begin{equation}
    c_{N,B}\simeq 6\times 10^{-4}
    \left(\frac{m_{N,B}}{5\gev}\right)
    \left(\frac{G_{F,B}}{G_{F,A}/10}\right).
\end{equation}
For a twin nucleus of mass number $\Anum$, we take the coherent coupling $\Anum\,c_{N,B}$.

\vspace{2cm} 

\bibliographystyle{utphys3}
\bibliography{Refs}

\end{document}